# Solitons supported by localized parametric gain

Fangwei Ye,[1] Changming Huang,[1] Yaroslav V. Kartashov,[2] and Boris A. Malomed[3]

[1]*Department of Physics, The State Key Laboratory on Fiber Optic Local Area, Communication Networks and Advanced Optical Communication Systems, Shanghai Jiao Tong University, Shanghai 200240, China*
[2]*Institute of Spectroscopy, Russian Academy of Sciences, Troitsk, Moscow Region, 142190, Russia*
[3]*Department of Physical Electronics, School of Electrical Engineering, Faculty of Engineering, Tel Aviv University, Tel Aviv 69978, Israel*
*Corresponding author: fangweiye@sjtu.edu.cn

We address the existence and properties of one-dimensional solitons maintained by localized parameter gain in focusing and defocusing lossy nonlinear media. Localized parametric gain supports both fundamental and multipole solitons. We found that the family of fundamental solitons is partly stable in focusing nonlinear medium, and completely stable in defocusing medium, while all higher-order solitons are unstable. In addition to numerical results, the existence threshold for the solitons, and a particular stable exact solution are obtained in an exact analytical form.

Spatial optical solitons appear when diffraction of light is balanced by a nonlinear response of the medium [1]. Optical dissipative solitons are self-trapped beams supported by the additional balance between losses and gain [2, 3]. In dissipative systems the uniform losses can be compensated by a spatially localized gain applied at a "hot spot", thus giving rise to dissipative solitons *pinned* to the hot spots, which were studied in diverse settings [4-11]. However, all previous settings, where solitons residing on hot spots were obtained, employ usual linear or nonlinear gain, which is not sensitive to the phase of the beam. In this Letter, we address the interplay between the uniform linear loss, Kerr nonlinearity, and local parametric gain applied at the hot spot, which can be induced as in usual parametric amplifiers [12-14]. The spatial localization of gain adds new features to properties of solitons, and it is the main difference from previous works on solitons supported by the parametric gain in uniform media [12, 15-20].

We thus consider the light propagation along the $\xi$ axis, governed by the nonlinear Schrödinger equation for the scaled field amplitude $q$:

$$i\frac{\partial q}{\partial \xi} = -\frac{1}{2}\frac{\partial^2 q}{\partial \eta^2} - (k+i\gamma)q + (\sigma - i\alpha)|q|^2 q - a e^{-\eta^2/d^2} q^*, \quad (1)$$

where the transverse and longitudinal coordinates, $\eta$ and $\xi$, are scaled to the beam's width $x_0$ and diffraction length $L_{\text{dif}} = 2\pi n x_0^2/\lambda$, respectively, $\gamma > 0$ and $\alpha \geq 0$ are strengths of the linear and cubic losses, $\sigma = -1 \; (+1)$ corresponds to the focusing (defocusing) Kerr nonlinearity, while $a$, $d$, and $k$ are the amplitude, spatial width, and detuning of the localized parametric gain, and the asterisk stands for the complex conjugate. Below, we consider the basic model with $k = \alpha = 0$ (i.e. no detuning and zero cubic losses), unless it is stated otherwise. For the beam with $x_0 = 40$ $\mu$m and wavelength $\lambda = 1.55$ $\mu$m, propagating in a material with refractive index $n \simeq 1.5$ and nonlinearity coefficient $n_2 = 3 \times 10^{-14}$ cm$^2$/W, $q \sim 1$ corresponds to the field intensity $\sim 1$ GW/cm$^2$, and $\gamma \sim 1$ implies the absorption coefficient $\sim 1.02$ cm$^{-1}$.

Stationary soliton solution to Eq. (1) were looked for as $q(\eta,\xi) = u(\eta)\exp[i\phi(\eta)]$, with real amplitude and phase. Figure 1 shows typical profiles of numerically generated fundamental (nodeless) solitary modes. Parametric gain can support spatially localized solitons in both focusing and defocusing media. With the increase of the gain, the profile of soliton in focusing medium changes from bell-shaped into multi-peaked one [compare Figs. 1(a)-1(c)]. In Fig. 1 we also plot the quantity $j(\eta) \equiv u^2 d\phi/d\eta$, which indicates the direction of internal currents inside soliton solutions (which are always directed from the domain with parametric gain into domain with linear losses). The antisymmetry of the current corroborates that the solitons are indeed an outcome of the balance between the loss and gain. In addition to the fundamental solitons, dipoles and higher-order modes exist too, although parametric gain is bell-shaped (see Fig. 2).

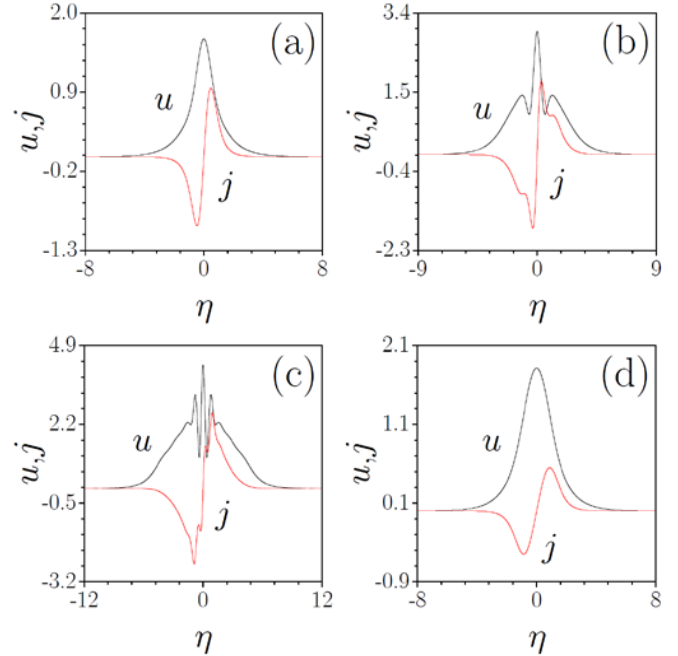

Fig. 1. (Color online) Distributions of the field's absolute value $u$ and current $j$ in fundamental solitons pinned to the hot spot in the focusing medium with (a) $a=2$, (b) $a=4$, and (c) $a=6$. (d) The same in the defocusing medium with $a=4$. The soliton in (c) is unstable, the others being stable. For the sake of the presentation, the current in panels (b) and (c) is multiplied by $0.3$ and $0.1$, respectively. These and all other examples are shown for $\gamma = 1$.

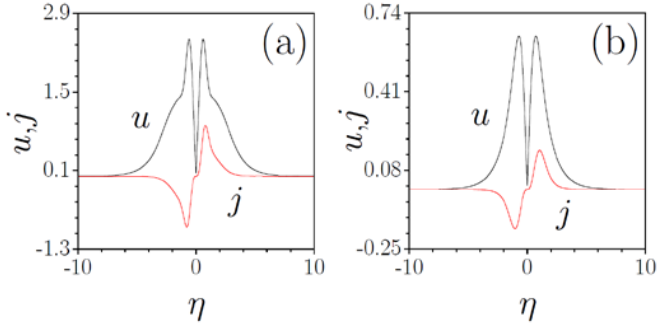

Fig. 2. The same as in Fig. 1, but for unstable dipole solitons in the focusing (a) and defocusing (b) media with $a=4$. The current in (a) is multiplied by $0.1$.

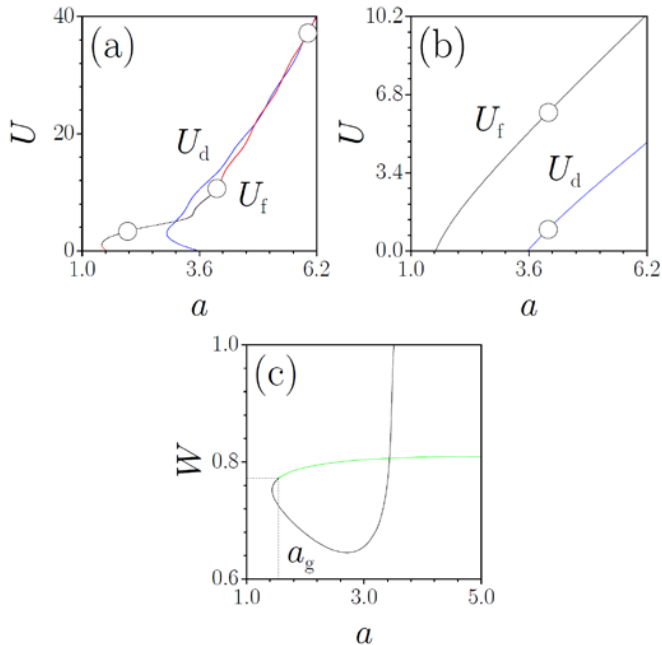

Fig. 3. (Color online) The energy flow versus the gain for fundamental ($U_\text{f}$) and dipole ($U_\text{d}$) solitons in the focusing (a) and defocusing (b) media. Stable and unstable parts of the fundamental-soliton branch are shown black and red, respectively. Dipoles are completely unstable. Circles in (a) and (b) correspond to the solitons shown in Figs. 1(a)-1(c) and Figs. 1(d), 2(b) respectively. (c) The integral width versus gain for fundamental solitons in the focusing and defocusing media (black and green curves, respectively). The point at the junction of the curves represents the gain-guided linear mode, existing at $a=a_\text{g}$.

Soliton properties are summarized in Fig. 3, where the dependences of energy flow $U = \int_{-\infty}^{+\infty} u^2 d\eta$ and integral width $W^2 = U^{-1}\int_{-\infty}^{+\infty} u^2 \eta^2 d\eta$ on gain amplitude $a$ are plotted, for both fundamental and dipole solitons. While in the defocusing medium $U$ is a monotonously growing function of the gain for both fundamental and dipole solitons [Fig. 3(b)], in focusing medium the initial short segment of the $U(a)$ dependence has a negative slope, turning forward at a threshold value of the gain, $a = a_\text{th}$. With further increase of $U$, the soliton in focusing medium develops additional peaks in its profile, which makes $U(a)$ dependence somewhat unsmooth [Fig. 3(a)].

In the limit of $U \to 0$ Figs. 3(a) and 3(b) demonstrate that the solitons in both focusing and defocusing media appear at the same initial value of the gain, $a = a_\text{g}$, which implies the existence of a linear localized "gain-guided" mode at this point. This fact is further reinforced by Fig. 3(c), which shows that the dependences of the soliton's width on the gain in the focusing and defocusing cases connect at $a = a_\text{g}$, corresponding to the linear-mode limit. The value $a_\text{g}$ and the respective linear mode can be found in an analytical form (for all values of $k$) if the narrow Gaussian in Eq. (1) is approximated by the $\delta$-function:

$$\exp(-\eta^2/d^2) \approx \pi^{1/2} d\delta(\eta) \qquad (2)$$

Then, $a_\text{g}$ is determined by the condition that linearized version of Eq. (1) admits an exact solution in the form of $q = \exp(i\theta - C|\eta|)$. After simple manipulations, one finds that this solution exists at

$$a_\text{g}^2(k) = (2/\pi d^2)(k^2 + \gamma^2)^{1/2}, \qquad (3)$$

with $4\theta = \tan^{-1}(\gamma/k)$, $C^2 = 2(k^2+\gamma^2)^{1/2}\exp(-4i\theta)$. Direct comparison demonstrates that Eq. (3) is very close to the numerical findings.

Under the same approximation (2), and setting $k=0$ as above, it is possible to find a particular *exact* soliton solution of Eq. (1) for $\sigma = +1$ (defocusing nonlinearity) and nonzero cubic losses, $\alpha = 3$:

$$q(\eta) = (\gamma/2)^{1/2}\exp(i\pi/8)[\sinh(\gamma^{1/2}(|\eta|+\kappa))]^{-1+i}, \quad (4)$$

where $\tanh(\gamma^{1/2}\kappa) = a_\text{g}(0)/a$, with $a_\text{g}(k)$ given by Eq. (3). This solution, whose energy flow is $U = (\pi/2)^{1/2}d[a-a_\text{g}(0)]$, is stable. A similar exact solution can be found for $k \neq 0$:

$$q(\eta) = (3\gamma/2\alpha)^{1/2}\exp(i\delta)[\sinh(\gamma^{1/2}(|\eta|+\kappa)/\mu^{1/2})]^{-1+i\mu}, \quad (5)$$

where $\mu = (9\sigma^2/4\alpha^2 + 2)^{1/2} - 3\sigma/2\alpha$, $2\delta = \arctan\mu$ and $\tanh^2(\gamma^{1/2}\kappa/\mu^{1/2}) = \gamma(\mu+\mu^{-1})/\pi a^2 d^2$. It exists at $k = (\gamma/2)(\mu - \mu^{-1})$ (at $k=0$, this constraint goes over into the above-mentioned one, $\alpha = 3$), which corresponds to the defocusing and focusing nonlinearity at $k < 2^{-3/2}\gamma$ and $k > 2^{-3/2}\gamma$.

The soliton stability was analyzed numerically, by taking perturbed solutions as $q(\eta,\xi) = [u\exp(i\phi) + f\exp(\delta\xi) + g^*\exp(-\delta^*\xi)]$, where $f$ and $g$ are perturbation modes with a complex growth rate, $\delta = \delta_r + i\delta_i$. The soliton is unstable if solutions with $\delta_r > 0$ can be found and completely stable otherwise. The substitution of perturbed field into Eq.(1) and subsequent linearization yields coupled linear equations for $f$ and $g$. The outcome of the analysis is that all fundamental solitons are stable under the defocusing nonlinearity. In the focusing medium, the initial negative-slope segment of the $U(a)$ branch for fundamental solitons is strongly unstable, while a significant stable portion exists on positive-slope segment above the turning point, in the interval $a_{\text{th}} \leq a \leq a_{\text{cr}}$. The fundamental solitons again become unstable above the critical value, $a = a_{\text{cr}}$. As expected, $a_g$, $a_{\text{cr}}$ and $a_{\text{th}}$ significantly depend on the coefficient of linear losses $\gamma$ [Fig.4(a)]. For $\gamma = 1$ one has $a_{\text{th}} \approx 1.4$ and $a_{\text{cr}} \approx 4.2$ (thus if $\gamma \sim 1$ cm$^{-1}$ in physical units, the minimum gain to compensate the loss is $a_{\text{th}} \approx 1.4$ cm$^{-1}$, while the critical gain is $a_{\text{cr}} \approx 4.2$ cm$^{-1}$). The dependence of perturbation growth rate on $a$ is displayed in Fig.4(b). All higher-order solitons were found to be unstable in both focusing and defocusing medium.

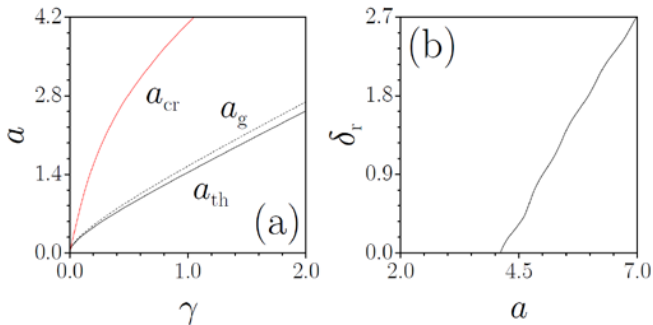

Fig. 4. (a) The critical, threshold, and initial (corresponding to the linear limit) values of the gain, $a_{\text{cr}}$, $a_{\text{th}}$, and $a_g$, versus linear loss $\gamma$ for fundamental solitons in the focusing medium. (b) The real part of the instability growth rate versus $a$ at $\gamma = 1$.

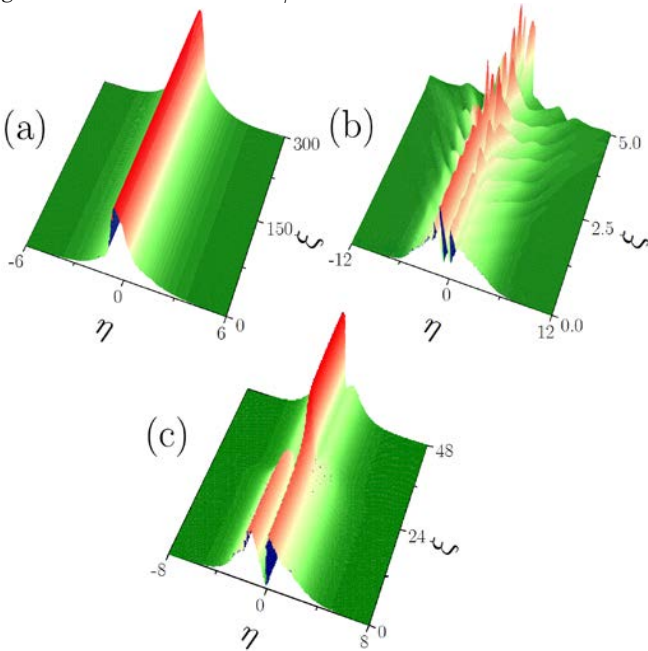

Fig. 5. (a) Stable (a) and unstable (b) propagation of perturbed fundamental solitons from the upper branch in Fig. 3(a) (in the focusing medium) at $a = 3$ and $a = 6$, respectively. (c) The transformation of an unstable dipole in the case of focusing medium into a stable fundamental soliton at $a = 3.5$. In all cases $\gamma = 1$.

The results of linear stability analysis are corroborated by direct simulations of Eq. (1) for slightly perturbed initial conditions. As expected, solitons which are predicted to be stable (in particular, all the fundamental solitons in the defocusing medium) maintain their original shapes [Fig. 5(a)]. Unstable fundamental solitons in the focusing medium quickly develop oscillations and usually relax into stable solitons [one example of unstable evolution is shown in Fig. 5(b)]. Dipole solitons may transform into stable fundamental solitons by shedding radiation waves [ Fig. 5(c)].

To summarize, we have studied solitons in Kerr media with the linear background loss and spatially localized parametric gain. The gain can balance the uniform loss and support different types of solitons, the family of the fundamental ones being completely stable under defocusing nonlinearity and partly stable in the case of focusing nonlinearity.